\documentclass[aps,pre, reprint, twocolumn,superscriptaddress]{revtex4}

\usepackage{epsfig}
\usepackage{bm}
\usepackage[retainorgcmds]{IEEEtrantools}
\usepackage{graphicx}
\usepackage{mathrsfs}
\usepackage{amsmath}
\usepackage{amssymb}

\begin{document}

\title{Fourier's law from a chain of coupled planar harmonic oscillators
under energy conserving noise}

\author{Gabriel T. Landi}
\affiliation{Centro de Ci\^encias Naturais e Humanas, Universidade Federal do ABC, Santo Andr\'e, S\~ao Paulo, 09210-170, Brazil}
\affiliation{Instituto de F\'{\i}sica, Universidade de S\~ao Paulo, Caixa Postal 66318, 05314-970, S\~ao Paulo, Brazil}
\author{M\'ario J. de Oliveira}
\affiliation{Instituto de F\'{\i}sica, Universidade de S\~ao Paulo, Caixa Postal 66318, 05314-970, S\~ao Paulo, Brazil}
\begin{abstract}

We study the transport of heat along a chain
of particles interacting through a harmonic potential
and subject to heat reservoirs at its ends. Each particle
has two degrees of freedom and is 
subject to a stochastic noise that produces infinitesimal changes in the velocity
while keeping the kinetic energy unchanged.
This is modelled by means of a Langevin equation with multiplicative noise. 
We show that the introduction of this
energy conserving stochastic noise leads to Fourier's law. 
By means of an approximate solution that becomes exact in the thermodynamic limit, 
we also show that the heat conductivity $\kappa$ behaves
as $\kappa = a L/(b+\lambda L)$ for large values of the intensity
$\lambda$ of the energy conserving noise and large chain sizes $L$. 
Hence, we conclude that in the thermodynamic limit the heat conductivity is finite and given by $\kappa=a/\lambda$.


\end{abstract}

\maketitle

%
\section{introduction}
%

Fourier's law of heat conduction states that the heat flux $J$ 
is proportional do the gradient of temperature, that is,
$J=-\kappa\nabla T$, where $\kappa$ is the heat conductivity. 
Since this law is understood as a {\it macroscopic} description of a 
non-equilibrium phenomena, it seems natural to address the problem of  deriving 
Fourier's law from a \emph{microscopic} model. However, this task has proved to be incredibly
challenging. Indeed, despite being over two hundred years old,
to this day no definitive microscopic model for this law has yet
been agreed on. The first attempt was made by Rieder, Lebowitz and Lieb
\cite{rieder67} who considered a linear chain of particles connected
by harmonic forces, with the first and last particles coupled to
Langevin reservoirs at different temperatures. Their calculations
showed that this model yields a ballistic (instead of a diffusive)
heat flow. If we write $J=\kappa\Delta T/L$, where $L$ is the size
of the system, then ballistic flow means that $J$ is constant so that,
in the thermodynamic limit ($L\to\infty$), $\kappa$ diverges. Hence,
the finiteness of $\kappa$ in the thermodynamic limit serves as a
criterion for the validity of Fourier's law.

The ballistic nature of the harmonic chain incite the idea that a new
ingredient is necessary to yield the correct diffusive behavior.
Indeed, several variations of the harmonic chain have been studied
in the past decades. These include the use of anharmonic interactions
\cite{lepri97,aoki01,eckmann04,cipriani05,pereira06,mai07,lukkarinen08,
gersch10,bernardin11,roy12,beijeren12},
systems with disorder \cite{dhar01,pereira08},
self-consistent reservoirs \cite{bolsterli70,bonetto04,pereira04,delfini06},
and many others \cite{narayan02,grassberger02,deutsch03,casati03,
basile09,lepri03,iacobucci10,dhar08,dubi09a,dubi09b,lepri09}.
Many of these attempts lead to anomalous diffusion,
for which $\kappa$ is also infinite. 
Some, however, do lead to
Fourier's law. An important example is the self-consistent reservoir
model introduced by Bolsterli, Rich and Visscher \cite{bolsterli70}.
In this model all
particles (and not just the first an the last) are connected to heat
reservoirs whose temperatures are chosen such that, in the steady
state there is no exchange of energy between the reservoirs and 
the inner particles of the chain (i.e., all except the first and the last).

An essential requirement in the construction of a microscopic model leading to Fourier's law 
is that heat should be exchanged only through the end points of the chain --  
no energy should enter or leave the system through the inner particles. Notice that the self-consistent model, strictly speaking,
does not meet this requirement. 
A  recent approach that fulfils this requirement and leads to Fourier's law (in the harmonic chain),
is based on the introduction of an energy conserving noise that flips
the sign of the velocity with some given rate \cite{landi13,dhar11}.
This noise models the interaction of the chain with additional degrees
of freedom in the medium.
In the present paper we are concerned with a new type of energy-conserving noise,
 which closely resembles elastic collisions in a solid and, as we will show, leads to Fourier's law.
This is accomplished by the introduction of infinitesimal random
changes of the velocity, modelled by a Langevin equation with multiplicative
noise devised so that it conserves the kinetic energy.

The main features of our study are as follows. 
First, it indicates that
the relevant property required to induce Fourier's law is the
energy-conserving nature of the noise and not its fine details or
the mechanism with which it is implemented. Second, when compared
to the aforementioned velocity-flipping model, this new noise has 
a more natural interpretation as elastic collisions
of the atoms in a crystal with other microscopic degrees of freedom.
Third, by modelling this noise by means of a Langevin equation with
multiplicative noise, it becomes possible to recast the problem in
terms of a system of linear equations for the position-velocity
covariances.
Solving numerically this linear problem is not only
faster than solving numerically the Langevin equation, but also
gives a much deeper insight into the problem. From the covariances
we obtain an approximate expression for the heat conductivity for large
chain sizes and large intensities of the energy-conserving noise. 
This expression, as will be shown, becomes exact in the thermodynamic limit. 
Moreover, we also present  exact expressions in the opposite situation of small system sizes.  
Finally, the nonequilibrium steady state (NESS) is shown to be Gaussian so that it is entirely
defined by the covariances.

We will consider the usual linear chain with harmonic potentials
and with the first and last particles connected to Langevin heat
baths at different temperatures. However, we allow each particle
to have two degrees of freedom. This simple variation enables us
to introduce {\it infinitesimal} random rotations of the velocities
of each particle. 
To see how this type of noise is introduced let us 
consider for the time being only a single particle with unit mass,
free to move in the $xy$ plane and let $v$ and $u$ denote the 
velocity components of this particle in the $x$ and $y$ directions respectively. 
Now, consider
the following Langevin equations with multiplicative noise \cite{lemos02}:
\begin{equation}
\frac{dv}{dt} = -\lambda v -\sqrt{2\lambda}\,u\, \zeta,
\label{1}
\end{equation} 
\begin{equation}
\frac{du}{dt} = -\lambda u +\sqrt{2\lambda}\,v\, \zeta,
\label{2}
\end{equation} 
where $\zeta(t)$ is a standard Gaussian white noise and
$\lambda$ represents the rate (or the intensity) of the noise;
notice that the noise $\zeta(t)$ is the same in both equations
but their signs are distinct. 
One can easily show that the magnitude of the velocity ($v^2+u^2)^{1/2}$
is invariant. From this result, it follows that the kinetic energy is
conserved so that these Langevin equations appropriately describe
random elastic collisions of the particle with the medium. 
They make up the key point of our model. For completeness, we also write the Fokker-Planck equation associated to the Langevin equations (\ref{1}) and (\ref{2}),
\begin{IEEEeqnarray}{rCl}
\frac{\partial P}{\partial t} &=& \lambda \left\{ \frac{\partial (vP)}{\partial v} + \frac{\partial (uP)}{\partial u} \right.\\[0.2cm]
&&\left.+ u^2 \frac{\partial^2P}{\partial v^2} + v^2 \frac{\partial^2 P}{\partial u^2} - 2 \frac{\partial^2 (uvP)}{\partial u \partial v} \right\}.\nonumber
\end{IEEEeqnarray}
It shows that the intensity of the collisions, $\lambda$, may be taken as a characteristic time constant. 

As it will be shown below, the inclusion of this new type of random elastic
collisions in the harmonic chain correctly leads to Fourier's law.
Moreover, in the thermodynamic limit, we find that $\lambda$
acts as a relevant parameter. That is, no matter how small
it is, as long as $\lambda\neq0$, the system will obey
the correct diffusive behaviour.
When $\lambda=0$, we recover the ballistic model of
Rieder, Lebowitz and Lieb \cite{rieder67}. 

When $\lambda$ and the system size $L$ are large enough,
it is possible to obtain an exact result for the heat conductivity
which, as we will show, behaves as
\begin{equation}\label{eq:kappa_primeira}
\kappa = \frac{aL}{b+\lambda L},
\end{equation}
where $a$ and $b$ are independent of $\lambda$ and $L$, even though they depend  on other parameters of the model. 
Therefore, in the thermodynamic
limit the heat conductivity is finite and given by $\kappa=a/\lambda$.

%
\section{Model}
%

We now describe the model studied in this paper. Consider a chain
of $L$ particles, each with two degrees of freedom. Their
positions are denoted by $x_i$ and $y_i$ and their velocities
by $v_i=dx_i/dt$ and $u_i=dy_i/dt$, with $i = 1,\ldots,L$. The equations of motions, assuming unit mass,  are 
\begin{equation}
\frac{dv_i}{dt} =  f_i -\lambda v_i - \sqrt{2\lambda}\, u_i\zeta_i
-\gamma_i v_i + \sqrt{2\gamma_i T_i}\,\xi_i^x,
\label{5}
\end{equation}
\begin{equation}
\frac{du_i}{dt} = g_i -\lambda u_i + \sqrt{2\lambda}\, v_i\zeta_i
-\gamma_i u_i + \sqrt{2\gamma_i T_i}\,\xi_i^y,
\label{6}
\end{equation}
where $f_i$ and $g_i$ are the $x$ and $y$ components of the force
acting on the $i$-th particle and
$\zeta_i(t)$, $\xi_i^x(t)$ and $\xi_i^y(t)$ are independent standard
Gaussian white noises. The parameters $\gamma_i$
are zero except when $i=1$ and $i=L$, in which case 
$\gamma_1=\gamma_L=\gamma$. They describe the contact of the system with two reservoirs at temperatures $T_1=T_A$ and   
$T_L= T_B$. The Boltzmann constant is set to unity. 
We note that the most relevant parameter is $\lambda$, the intensity
of the random elastic collisions.

The set of Langevin equations (\ref{5}) and (\ref{6}) may also
be interpreted as describing two coupled one-dimensional chains of particles.
One described
by the variables $x_i$ and $v_i$ and the other by the variables
$y_i$ and $u_i$. The energy-conserving noise is interpreted as
a stochastic noise that changes the velocity of two particles
belonging to distinct chains in such a way that their combined
kinetic energies remain constant. This interpretation is very
natural and can be extended, for instance, to several one-dimensional
chains.

The Fokker-Planck equation associated to the Langevin equations
(\ref{5}) and (\ref{6}), which describes the time evolution of the
probability distribution, is given by
\begin{IEEEeqnarray}{rCl}
\frac{\partial P}{\partial t} &=& -\sum_i \left(\frac{\partial v_iP}{\partial x_i}
+\frac{\partial u_iP}{\partial y_i}
+\frac{\partial \hat{f}_i P}{\partial v_i}
+\frac{\partial \hat{g}_i P}{\partial u_i}
\right)\nonumber \\[0.2cm]
&&+\sum_i\left(
 \frac{\partial^2 D_i^x P}{\partial v_i^2}
+\frac{\partial^2 D_i^y P}{\partial u_i^2}
-2\lambda\frac{\partial^2 v_iu_iP}{\partial v_i\partial u_i}
\right),
\label{9}
\end{IEEEeqnarray}
where
\begin{equation}
\hat{f}_i = f_i - (\gamma_i + \lambda) v_i,
\qquad
\hat{g}_i = g_i - (\gamma_i + \lambda) u_i,
\end{equation}
\begin{equation}
D_i^x = \gamma_i T_i +\lambda u_i^2,
\qquad
D_i^y = \gamma_i T_i +\lambda v_i^2.
\end{equation}

The forces are assumed to be conservative, that is, they
are the gradient of a potential energy $U$, 
$f_i=-\partial U/\partial x_i$ and $g_i=-\partial U/\partial y_i$.
When the system is uncoupled to the heat reservoirs, the total
energy
\begin{equation}
E = \sum_{i=1}^L\frac{m}{2}(v_i^2+u_i^2) + U
\end{equation}
is a constant of motion. Thus, in this case the system  evolves in isolation and,
due to the random elastic collisions, is ergodic; that is,
it reaches an equilibrium given by the Gibbs microcanonical 
distribution. When the system is coupled to the heat baths the
change in the total energy is entirely due to the exchange of
energy with the heat bath. If the temperatures of the heat baths
are the same, the equilibrium distribution is the Gibbs canonical
distribution.

In this paper we focus on harmonic potentials, which 
yield closed equations for the covariances, as we shall see below. 
The harmonic potential $U$ that we use has the general form
\begin{equation}
U = \frac{1}{2}\sum_{ij}A_{ij}x_ix_j
+ \frac{1}{2}\sum_{ij}B_{ij}y_iy_j + \sum_{ij}C_{ij}x_iy_j,
\label{10}
\end{equation}
where $A_{ij}$, $B_{ij}$ and $C_{ij}$ are understood as the elements
of $L\times L$ matrices $A$, $B$ and $C$.

We have used several types of harmonic potentials and all lead
to Fourier's law. For definiteness, we shall consider here three
specific forms of $U$, all involving nearest-neighbor
interactions.

I) The first type of potential is  symmetric and uncoupled in $x$ and $y$.
It is given by
\begin{equation}
U_1 = \frac{k}{2}\sum_{i=0}^L [(x_i-x_{i+1})^2+(y_i-y_{i+1})^2].
\label{14}
\end{equation}
where $x_0= x_{L+1}=y_0 = y_{L+1}=0$. When compared to (\ref{10}),
we see that  $A$ is the tridiagonal matrix 
\begin{equation}
A = k \left(
\begin{array}{rrrrrrrr}
 2  		& 	-1 	&  	 0 	& 	0 	& 	0	&	\ldots  	& 	0  	&	0\\
-1 		&  	2  	&  	-1  	&  	0  	& 	0 	&  	\ldots  	&    	0 	&	0\\
 0 	 	& 	-1  	&	  2  	& 	-1  	&  	0 	&  	\ldots  	&   	0 	&	0\\
 0 	 	&  	0  	&  	-1  	&  	2  	& 	-1 	&  	\ldots  	& 	0 	&	0\\
  \vdots 	&  \vdots 	&  \vdots 	&  \vdots 	& \vdots  	&  	\ddots	& 	\vdots	&	\vdots\\
   0  		&  	0  	&   	0  	& 	0  	&  	0 	& 	-1 		&  	2	&	-1\\
 0  		&  	0 	&   	0  	& 	0  	&  	0 	&	 0 		&  	-1	&	2\\
\end{array}
\right),
\label{15}
\end{equation}
whereas $B=A$ and $C=0$. 
This choice of potential treats $x$ and $y$ on equal footing and does
not couple them. Hence, they are connected only through the
energy-conserving noise. 
In the stationary state the heat flux is determined by
the position-velocity covariance which, in this case, is given by
\begin{equation}
J = 2 k\langle x_i v_{i+1}\rangle.
\label{heat_flux1}
\end{equation}

II) The second type of potential is still symmetric in $x$ and $y$, but couples both directions. 
It is given by
\begin{IEEEeqnarray}{rCl}
U_2 &=& \frac{k}{2}\sum_{i=0}^L \Big[(x_i-x_{i+1})^2+(y_i-y_{i+1})^2 \nonumber \\
&&+2\alpha (x_i-x_{i+1})(y_i-y_{i+1})\Big],
\label{16}
\end{IEEEeqnarray}
where, again, $x_0= x_{L+1}=y_0 = y_{L+1}=0$.
The parameter $\alpha$ is chosen within the interval $0\leq\alpha\leq1$
in order to guarantee  mechanical stability.
Referring to equation (\ref{10}), we have $B=A$ and $C=2\alpha A$,
where $A$ is the tridiagonal matrix given by (\ref{15}).
In the stationary state the heat flux is given by
\begin{equation}
J = 2 k\Big[\langle x_i v_{i+1}\rangle + \alpha \langle x_i u_{i+1}\rangle\Big].
\end{equation}

III) The third type of potential is asymmetric and pinned in $y$.
It is given by
\begin{equation}
U_3 = \frac{k}{2}\sum_{i=0}^L (x_i-x_{i+1})^2 + 
\frac{k'}{2}\sum_{i=1}^L y_i^2.
\label{17}
\end{equation}
Now we have $x_0= x_{L+1}=0$.
In this case $C=0$, $A$ is the tridiagonal matrix given by (\ref{15})
and $B = (k'/2) I$ where $I$ is the $L\times L$ identity matrix. 
In the stationary state the heat flux is given by
\begin{equation}
J =  k\langle x_i v_{i+1}\rangle.
\end{equation}

Finally, in all cases the heat conductivity is computed from 
\begin{equation}
\kappa=\left| JL/\Delta T\right|.
\end{equation}

%
\section{Covariances}
%

%
\subsection{General harmonic potentials}
%

The linearity of the harmonic forces and the
type of energy-conserving noise we use here allow us
to find closed equations for the covariances, which can be
solved by standard (numerically exact) procedures.
It is useful to define 
$x=(x_1,\ldots,x_L)$, $v=(v_1,\ldots,v_L)$,
$y=(y_1,\ldots,y_L)$, and $u=(u_1,\ldots,u_L)$, all interpreted as column vectors. 
The $L\times L$ covariance matrices are defined by the expectation of the outer products:
\begin{equation}
X_1=\langle x x^\dagger\rangle, \qquad
X_2=\langle y y^\dagger\rangle, \qquad
X_3=\langle x y^\dagger\rangle,
\end{equation}
\begin{equation}
Y_1=\langle v v^\dagger\rangle, \qquad
Y_2=\langle u u^\dagger\rangle, \qquad
Y_3=\langle v u^\dagger\rangle,
\end{equation}
\begin{equation}
Z_1=\langle xv^\dagger\rangle, \qquad
Z_2=\langle yu^\dagger\rangle, 
\end{equation}
\begin{equation}
Z_3=\langle xu^\dagger\rangle, \qquad
Z_4=\langle yv^\dagger\rangle,
\end{equation}
The full $4L\times4L$ covariance matrix is
\begin{equation}
\Theta = \left(
\begin{array}{ll}
\Theta_1 & \Theta_3 \\[0.2cm]
\Theta_3^\dagger & \Theta_2 
\end{array}
\right) = \left(
\begin{array}{rrrr}
 X_1          &  Z_1            &  X_3          &  Z_3   \\[0.1cm]
 Z_1^\dagger  &  Y_1            &  Z_4^\dagger  &  Y_3   \\[0.1cm]
 X_3^\dagger  &  Z_4            &  X_2          &  Z_2   \\[0.1cm]
 Z_3^\dagger  &  Y_3^\dagger    &  Z_2^\dagger  &  Y_2   
\end{array}
\right).
\label{20}
\end{equation}

The evolution equations for the covariances are obtained from
the Fokker-Planck equation as follows. Consider for instance
the covariance $\langle x_ix_j\rangle$, which is an entry of $X_1$. Multiply both sides of
equation (\ref{9}) by $x_ix_j$ and take the average. The left-hand
side gives the time derivative $d\langle x_ix_j\rangle/dt$.
Performing the integrals in the right-hand side by parts,
as many time as necessary, we get the desired time evolution equation. 
Repeating this procedure for all covariances we reach the equation
\begin{equation}
\frac{d}{d t }\Theta = -(\Phi\Theta + \Theta\Phi^\dagger) + \Upsilon 
-\lambda \Psi,
\label{25}
\end{equation}
where the $4L\times4L$ matrix $\Phi$ is
\begin{equation}
\Phi = \left(
\begin{array}{ll}
\Phi_1 & \Phi_3 \\[0.2cm]
\Phi_3 & \Phi_2 \\
\end{array}
\right) = \left(
\begin{array}{rrrr}
 0   &  -I      &  0  &  0      \\
 A   &  \Gamma  &  C  &  0      \\
 0   &  0       &  0  &  -I     \\
 C   &  0       &  B  &  \Gamma \\
\end{array}
\right),
\label{26}
\end{equation}
where $I$ is the $L\times L$ identity matrix and $\Gamma$ is the diagonal
matrix with elements $\Gamma_{11} = \Gamma_{LL} = \gamma$, with all other entries being zero. The other $4L\times4L$
matrices appearing in equation (\ref{25}) are as follows:
\begin{equation}
\Upsilon = \left(
\begin{array}{ll}
\Upsilon_1 & 0 \\[0.2cm]
0          & \Upsilon_1 \\
\end{array}
\right) = \left(
\begin{array}{rrrr}
 0   &  0   &  0  &  0 \\
 0   &  D   &  0  &  0 \\
 0   &  0   &  0  &  0 \\
 0   &  0   &  0  &  D \\
\end{array}
\right),
\label{27}
\end{equation}
where $D$ is a $L\times L$ diagonal matrix with elements $D_{11}=2\gamma T_A$ and $D_{LL} = 2\gamma T_B$, again with all other entries zero. Moreover,
\begin{IEEEeqnarray}{rCl}
\Psi &=& \left(
\begin{array}{rr}
\Psi_1         &  \Psi_3 \\[0.2cm]
\Psi_3^\dagger &  \Psi_2 \\
\end{array}
\right)\nonumber \\[0.2cm]
&=&
\left(
\begin{array}{cccc}
 0           &  Z_1                      &  0           &  Z_3 \\[0.2cm]
 Z_1^\dagger &  2(Y_1-\bar{Y}_2)         &  Z_4^\dagger &  2(Y_3+\bar{Y}_3) \\[0.2cm]
 0           &  Z_4                      &  0           &  Z_2 \\[0.2cm]
 Z_3^\dagger &  2(Y_3^\dagger+\bar{Y}_3) &  Z_2^\dagger &  2(Y_2-\bar{Y}_1) \\[0.2cm]
\end{array}
\right),
\label{29}
\end{IEEEeqnarray}
where $\bar{Y}_1$, $\bar{Y}_2$ and $\bar{Y}_3$ are $L\times L$ diagonal matrices
composed by the diagonal elements of $Y_1$, $Y_2$ and $Y_3$, respectively.

In the stationary state, which interests us here, equation 
(\ref{25}) becomes
\begin{equation}
(\Phi\Theta + \Theta\Phi^\dagger) + \lambda \Psi = \Upsilon,
\label{25n}
\end{equation}
which can be written in an equivalent form,
in terms of $2L\times2L$ matrices,
\begin{equation}
(\Phi_1\Theta_1 + \Theta_1\Phi_1^\dagger) 
+(\Phi_3\Theta_3^\dagger + \Theta_3\Phi_3^\dagger) 
+\lambda \Psi_1 = \Upsilon_1,
\label{25a}
\end{equation}
\begin{equation}
(\Phi_2\Theta_2 + \Theta_2\Phi_2^\dagger) 
+(\Phi_3\Theta_3 + \Theta_3^\dagger\Phi_3^\dagger) 
+\lambda \Psi_2 = \Upsilon_1 ,
\label{25b}
\end{equation}
\begin{equation}
(\Phi_1\Theta_3 + \Theta_3\Phi_2^\dagger) 
+(\Phi_3\Theta_2 + \Theta_1\Phi_3^\dagger) + \lambda \Psi_3 = 0.
\label{25c}
\end{equation}
Note that Eqs.~(\ref{25a}) and (\ref{25b}) are coupled through the matrices $\Psi_1$ and $\Psi_2$, since in $\Psi_1$ there is a term containing $\bar{Y}_2$ and vice-versa [cf. Eq.~(\ref{29})]. 

Let us consider particular cases of these equations. 
When the potential is symmetric under the transformations
$x_i\rightleftharpoons y_i$ and $v_i\rightleftharpoons u_i$,
like that given by (\ref{14}) and (\ref{16}),  then $B=A$ so that $\Phi_2=\Phi_1$. Moreover, the Fokker-Planck equation 
will also be invariant under $x_i \rightleftharpoons y_i$ and $v_i \rightleftharpoons u_i$,
and so will the covariances,  leading to
the symmetric solution 
$\Theta_2=\Theta_1$, $\Psi_2=\Psi_1$ and $\Theta_3^\dagger=\Theta_3$.
Equations (\ref{25a})-(\ref{25c}) are then reduced to 
\begin{equation}
(\Phi_1\Theta_1 + \Theta_1\Phi_1^\dagger) 
+(\Phi_3\Theta_3 + \Theta_3\Phi_3) + \lambda \Psi_1 = \Upsilon_1,
\label{26a}
\end{equation}
\begin{equation}
(\Phi_1\Theta_3 + \Theta_3\Phi_1^\dagger) 
+(\Phi_3\Theta_1 + \Theta_1\Phi_3^\dagger) + \lambda \Psi_3 = 0.
\label{26c}
\end{equation}

If, furthermore, the variables $x$ and $y$
are not coupled, for instance when $U$ is given by (\ref{14}), then $C=0$
so that $\Phi_3=0$. In this case the equations (\ref{26a}) 
and (\ref{26c}) become two independent equations
for $\Theta_1$ and $\Theta_3$,
\begin{equation}
(\Phi_1\Theta_1 + \Theta_1\Phi_1^\dagger) 
+\lambda \Psi_1 = \Upsilon_1,
\label{30}
\end{equation}
\begin{equation}
(\Phi_1\Theta_3 + \Theta_3\Phi_1^\dagger) + \lambda \Psi_3 = 0.
\end{equation}
From this last equation, it follows that the interchain covariances vanish,
$\Theta_3=0$ and $\Psi_3=0$, and we are left only with equation~(\ref{30})  for $\Theta_1$.

Let us consider now an unsymmetrical potential like the one given
by (\ref{17}) for which $C=0$ so that
$\Phi_3=0$,   $A\neq0$ and $B\neq 0$. Moreover, $B$ is a diagonal matrix. 
In this case we get
\begin{equation}
(\Phi_1\Theta_1 + \Theta_1\Phi_1^\dagger)  
+\lambda \Psi_1 = \Upsilon_1,
\label{32a}
\end{equation}
\begin{equation}
(\Phi_2\Theta_2 + \Theta_2\Phi_2^\dagger) 
+\lambda \Psi_2 = \Upsilon_1 ,
\label{32b}
\end{equation}
\begin{equation}
(\Phi_1\Theta_3 + \Theta_3\Phi_2^\dagger) + \lambda \Psi_3 = 0.
\label{32c}
\end{equation}
The equation for $\Theta_3$ again gives $\Theta_3 =0$. The equations (\ref{32a})
and (\ref{32b}) are coupled through the diagonal covariances
$\bar{Y}_1$ and $\bar{Y}_2$ that appear in $\Psi_2$ and $\Psi_1$
respectively.

%
\subsection{Numerical results}
%

Before continuing with the analytical development of our model, we briefly stop to present a numerical analysis. 
Most of our discussion will focus on the symmetric potential $U_1$ in Eq.~(\ref{14}).  The other choices of potential do not change any of the important conclusions we shall obtain. 
In what follows we fix $k = 1$, $\gamma = 1$, $T_A = 1$ and $T_B = 2$. The free parameters are $\lambda$ (the intensity of the elastic collisions) and $L$ (the size of the system). For this choice of potential we may obtain the steady-state covariances by solving Eq.~(\ref{30}) numerically, which is simpler than the general Eq.~(\ref{25n}) valid for arbitrary harmonic potentials.  We then compute the heat flux from Eq.~(\ref{heat_flux1}) and finally the heat conductivity from the relation $\kappa = |JL/\Delta T|$.

\begin{figure}[!t]
\centering
\includegraphics[width=0.45\textwidth]{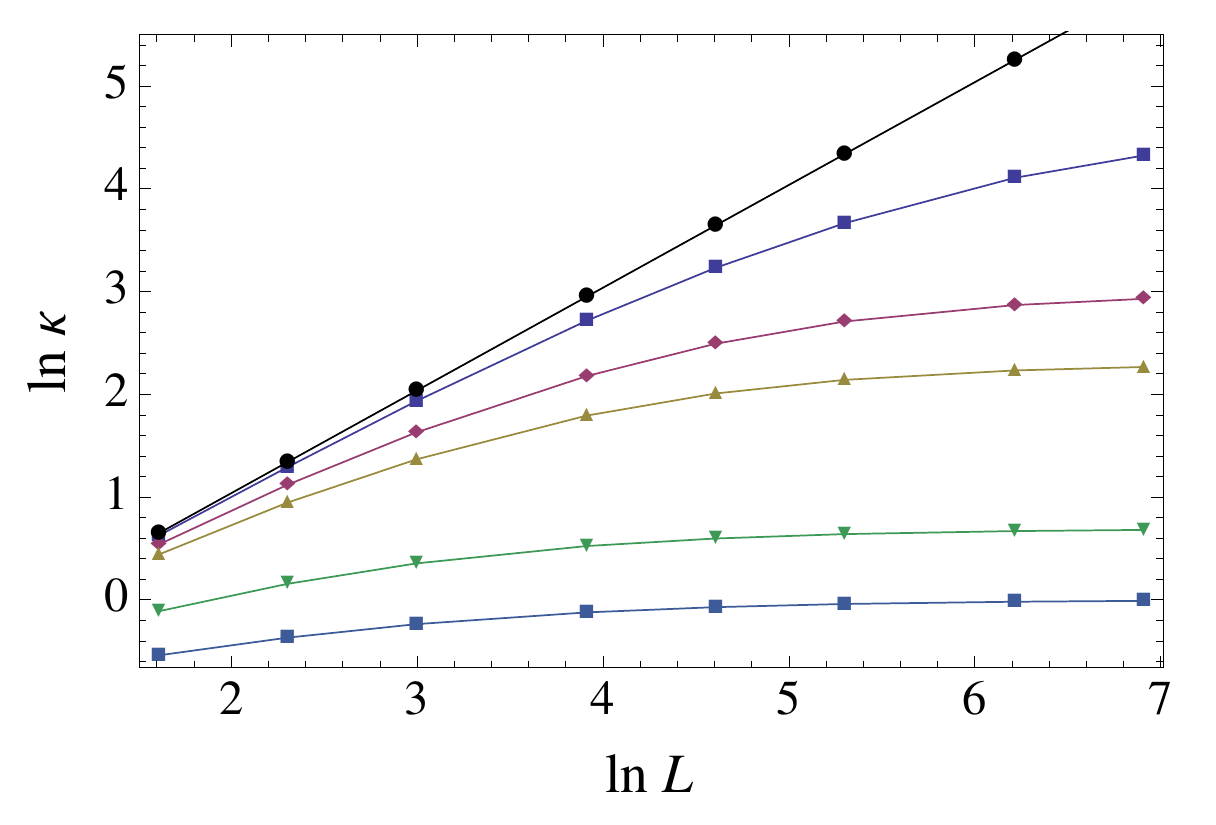}
\caption{\label{fig:kappa_L}Thermal conductivity $\kappa$ as a function of the system size $L$ for different values of $\lambda$, the intensity of the elastic collisions: from top to bottom, $\lambda = 0$, 0.01, 0.05, 0.1, 0.5 and 1. The calculations are for the potential $U_1$ in Eq.~(\ref{14}) with fixed $k=1$, $\gamma =1$, $T_A = 1$ and $T_B = 2$.}
\end{figure}

In Fig.~\ref{fig:kappa_L} we show results for $\kappa$ as a function of $L$, for both $\lambda = 0$ and $\lambda \neq 0$ (several values). When $\lambda = 0$ we see clearly that $\kappa \propto L$, which means that we recover the ballistic results of Ref.~\cite{rieder67}. In fact, these results can even be compared with their exact solution. This is so	 because, due to our choice of potential, when $\lambda = 0$ the $x$ and $y$ directions are independent, so that the heat conductivity is simply twice the original result for the one-dimensional chain. When $\lambda\neq 0$ we find that as $L$ increases, $\kappa$ tends  to a finite value. The rapidity with which this asymptotic limit is reached increases with increasing $\lambda$. Notwithstanding, we may conjecture that irrespective of how small $\lambda$ is, in the thermodynamic limit ($L\to\infty$) this  asymptotic value is always reached. This seems reasonable from the results of Fig.~\ref{fig:kappa_L} and will also be corroborated by further arguments to be given below. 
Fig.~\ref{fig:kappa_lambda} illustrates the dependence of $\kappa$ on $\lambda$ for different values of $L$. Note the broad range covered by $\lambda$, from $10^{-4}$ to $10^{2}$. This is a consequence of the efficiency of the numerical method just discussed. Fig.~\ref{fig:kappa_lambda} shows that, when $L\to \infty$, $\kappa \propto 1/\lambda$. 

\begin{figure}[!t]
\centering
\includegraphics[width=0.45\textwidth]{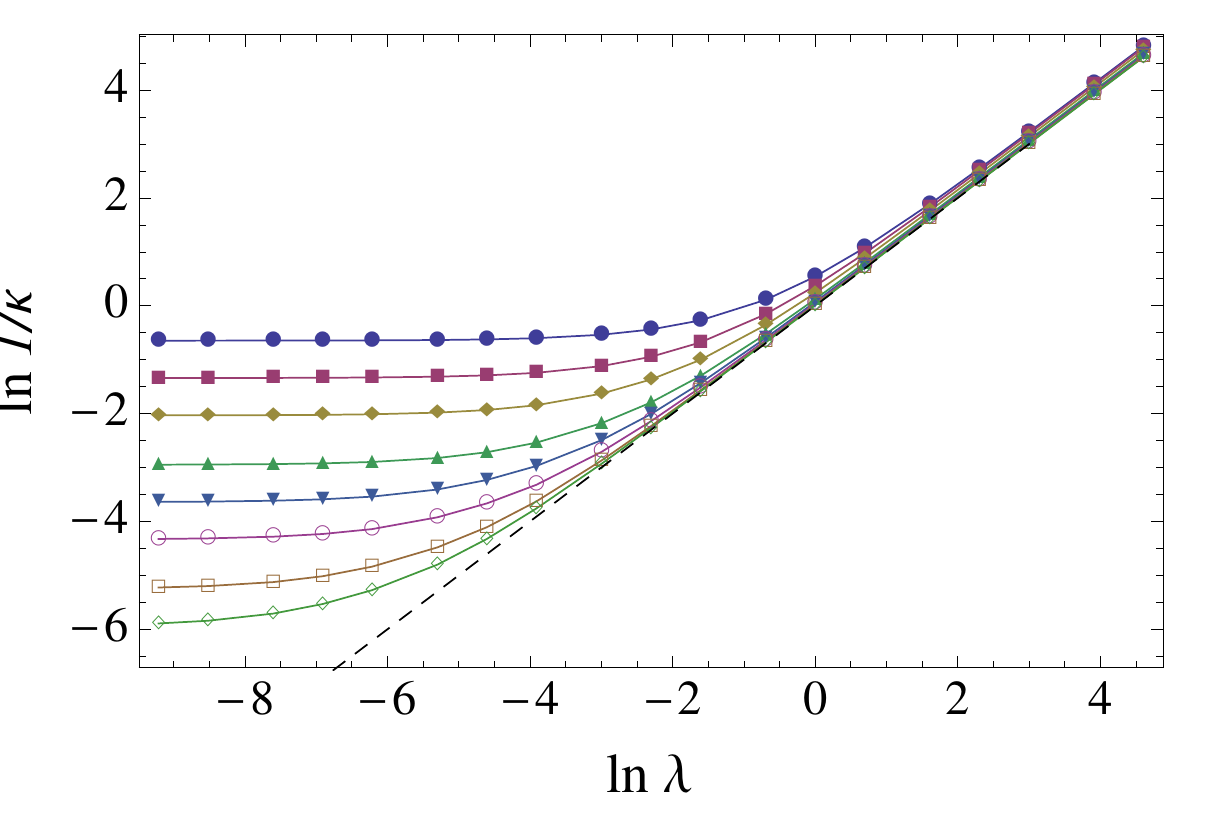}
\caption{\label{fig:kappa_lambda}$1/\kappa$~vs.~$\lambda$ for different values of $L$: from top to bottom $L = 5$, 10, 20, 50, 100, 200, 500 and 1000. The calculations are for the potential $U_1$ in Eq.~(\ref{14}) with the same parameters as in Fig~\ref{fig:kappa_L}.}
\end{figure}

\begin{figure}[!t]
\centering
\includegraphics[width=0.45\textwidth]{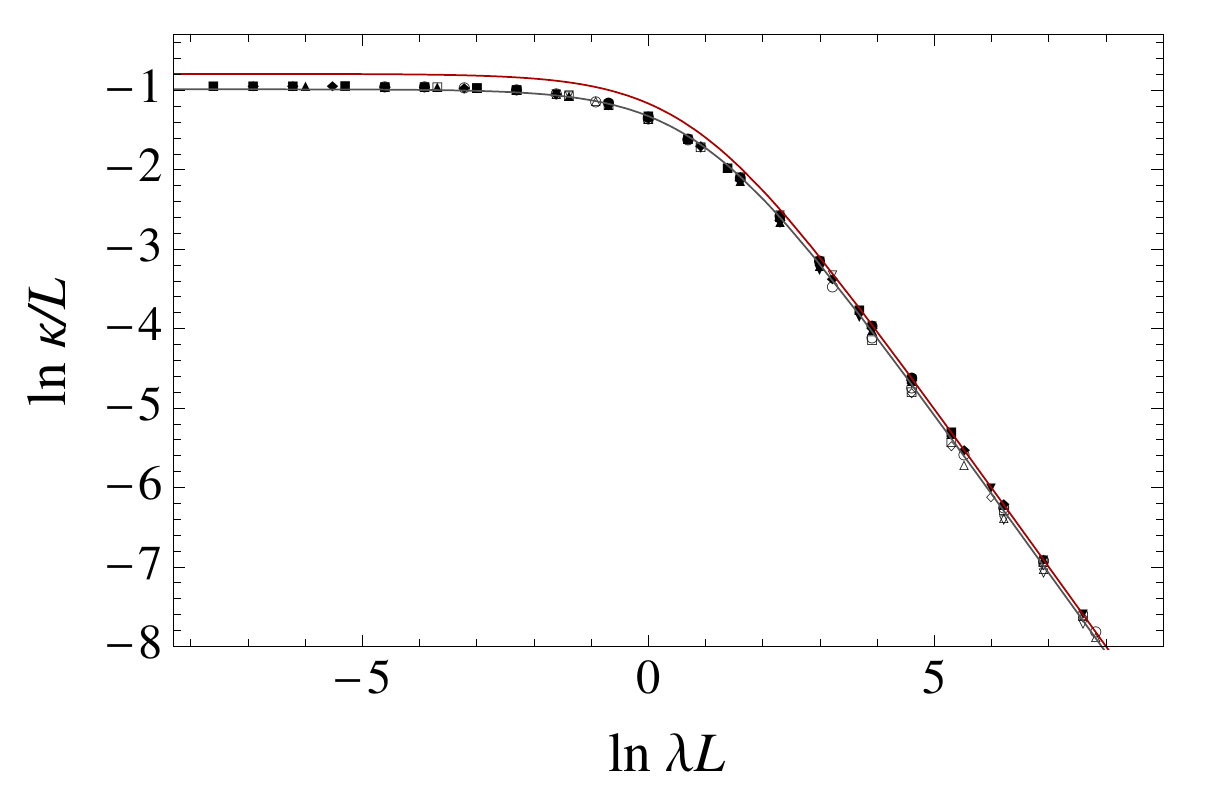}
\caption{\label{fig:FSC}Finite size scaling of $\kappa/L$~vs.~$\lambda L$ for the potential $U_1$, Eq.~(\ref{14}), with parameters $k=1$, $\gamma =1$, $T_A = 1$ and $T_B = 2$. The lower continuous line is a fitting
to the data points from Eq.~(\ref{eq:FSC}) with parameters $a'$ and $b'$. The upper continuous line represents the solution~(\ref{eq:kappa_inf}).}
\end{figure}

In summary, from Figs.~\ref{fig:kappa_L} and \ref{fig:kappa_lambda} we find the following scaling behaviour: when $L\to \infty$, $\kappa \propto 1/\lambda$ and when $\lambda = 0$, $\kappa \propto L$. We therefore assume the following scaling law \cite{landi13}:
\begin{equation}\label{eq:FSC}
\kappa = \frac{a' L}{b'+\lambda L},
\end{equation}
valid for small values of $\lambda$ and large values of $L$.
A fitting of this finite-size scaling is presented in Fig.~\ref{fig:FSC} where the collapse of the data points can be clearly observed. The finite-size scaling formula~(\ref{eq:FSC}) clearly shows that  $\lambda$ is a relevant parameter: as long as $\lambda \neq 0$, in the thermodynamic limit we always obtain a finite value of $\kappa$.  

\begin{figure}[!t]
\centering
\includegraphics[width=0.45\textwidth]{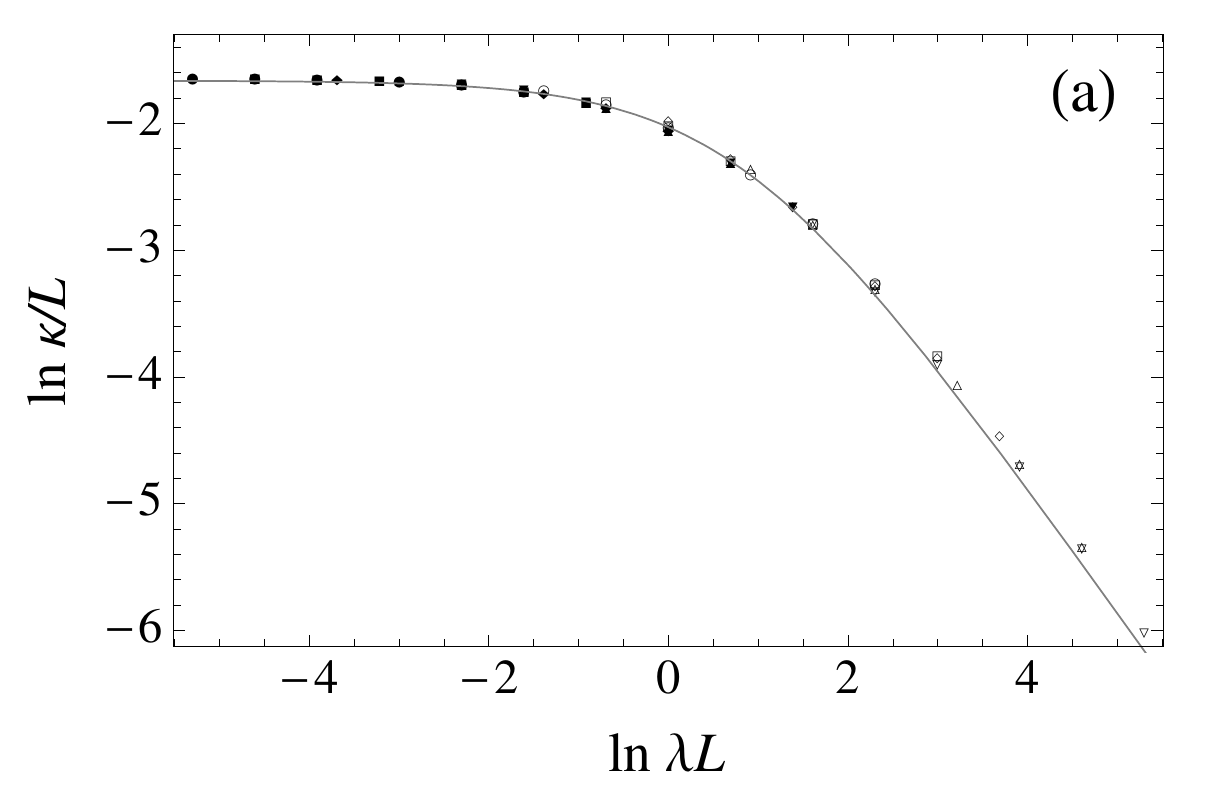}\\
\includegraphics[width=0.45\textwidth]{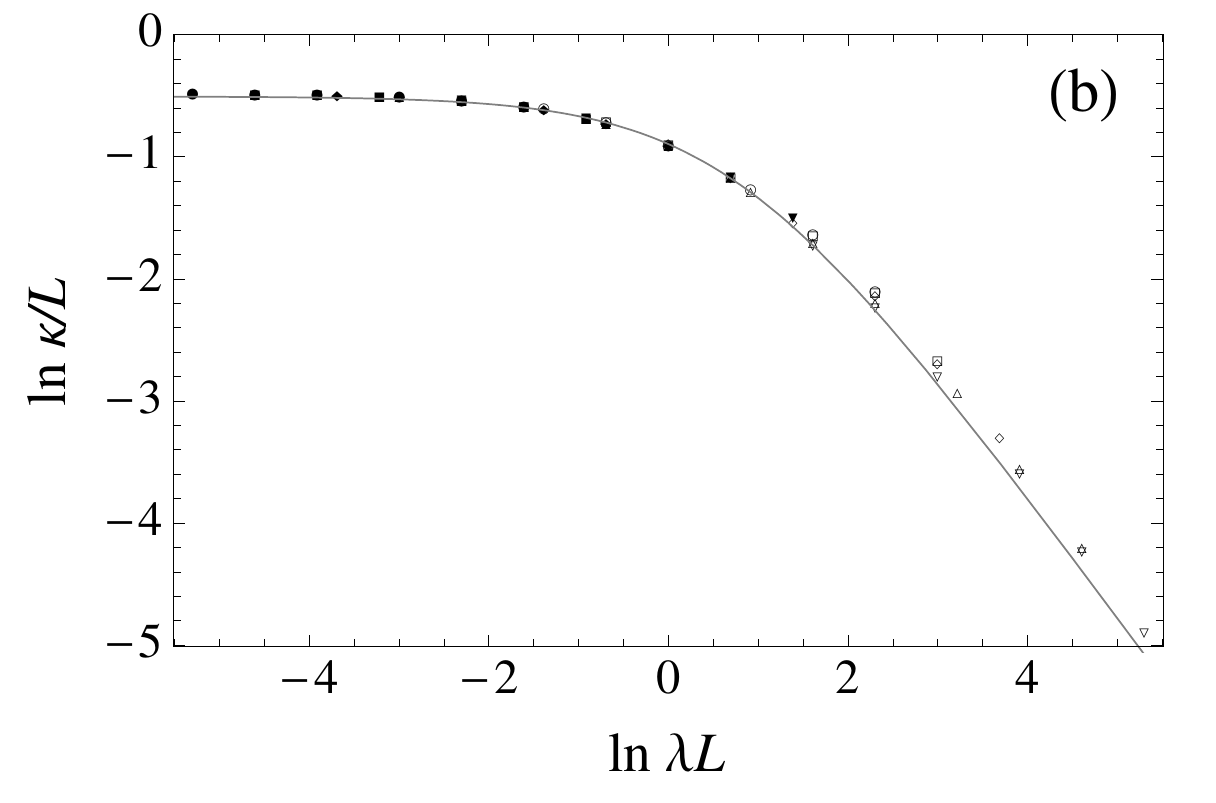} 
\caption{\label{fig:FSC2}Finite size scaling of $\kappa/L$~vs.~$\lambda L$ for the potentials (a) $U_2$ (with $\alpha = 0.5$) and (b) $U_3$ (with $k' = 1$) in Eqs.~(\ref{16}) and (\ref{17}). The solid line represents a fit from Eq.~(\ref{eq:FSC}) with fit parameters $a'$ and $b'$. The other parameters are $k=1$, $\gamma =1$, $T_A = 1$ and $T_B = 2$. }
\end{figure}

For completeness, in Fig.~\ref{fig:FSC2} we also present the scaling behaviour obtained for the other potentials, $U_2$ and $U_3$, defined  in Eqs.~(\ref{16}) and (\ref{17}) respectively. The parameters $a'$ and $b'$ in Eq.~(\ref{eq:FSC}) were fitted to the data. As can be seen, a very similar behaviour is obtained, which corroborates our claim that the choice of potential is unimportant in obtaining Fourier's law.

%
\section{\label{sec:ana}Analytical results}
%

%
\subsection{Symmetric and uncoupled potential}
%

We now return to Eqs.~(\ref{30}) for the covariances under the potential $U_1$ and show how it can be simplified.
Written explicitly, equation (\ref{30}) gives $Z^\dagger=-Z$ and
\begin{IEEEeqnarray}{rCl}
\label{33a}(AZ-ZA)+(\Gamma Y+Y\Gamma) + 2\lambda(Y-\bar{Y}) &=& D, \\[0.2cm]
\label{33b}(A X -X A) - (Z\Gamma + \Gamma Z) &=&  2\lambda Z, \\[0.2cm]
\label{33c}2Y - (X A + A X) - (Z\Gamma - \Gamma Z) &=&0,
\end{IEEEeqnarray}
where we have dropped the indices in $X_1$, $Y_1$ and $Z_1$.

Here we reach a remarkable result. 
These equations are exactly the same equations for the covariances in the
velocity-flipping model, equation (7) of reference \cite{dhar11}, which may therefore be interpreted as a particular case of our velocity-rotation model.
It is important to note, however, that the fact that the equations
for the covariances coincide does not imply that both models
are identical. For instance, the equations governing the evolution of the probability distribution of both models are entirely different,
which can be seen by noting that  in the present model it is described by a standard Fokker-Planck
equation, whereas in the velocity-flipping model the conserving noise
is modelled by a master equation-type term \cite{dhar11}.
The fact that both models give the same equations for the covariances and, hence, that both lead to Fourier's law, 
means that the rather sharp nature of the velocity-flipping model \cite{landi13,dhar11}
is not the relevant ingredient to induce Fourier's law.
What is in fact relevant is the energy-conserving nature of the noise.

We begin our analysis by subtracting the equilibrium solution $X^e$, $Y^e$ and $Z^e$ from the covariance matrices $X$, $Y$ and $Z$. Recall that the non-vanishing elements of $D$ are $D_{11}=2\gamma T_A$ and $D_{LL}=2\gamma T_B$.
The equilibrium covariances $X^{\rm e}$ and $Y^{\rm e}$ are solutions of
\begin{IEEEeqnarray}{rCl}
\label{34a}(\Gamma Y^{\rm e}+Y^{\rm e}\Gamma) &=& D^0, \\[0.2cm]
\label{34b}(A X^{\rm e} -X^{\rm e} A) &=&  0, \\[0.2cm]
\label{34c}2Y^{\rm e} - (X^{\rm e} A + A X^{\rm e}) &=&0,
\end{IEEEeqnarray}
where $D^0$ is the $L\times L$ diagonal matrix
with nonvanishing elements $D^0_{11}=D^0_{LL}=2\gamma T$
and $T=(T_A+T_B)/2$.
Notice that the velocity-velocity covariance matrix  $Y^e$ is 
diagonal and the position-velocity covariances vanish, $Z^e=0$.

Next we define the dimensionless matrices $X^*$, $Y^*$ and $Z^*$ by 
\begin{IEEEeqnarray}{rCl}
X &=& X^e + \frac{X^* \Delta T}{\gamma^2}, \\[0.2cm]
Y &=& Y^e + Y^* \Delta T,\\[0.2cm]
Z &=& Z^e + \frac{Z^* \Delta T}{\lambda},
\end{IEEEeqnarray}
 where $\Delta T = T_B-T_A$. The equations for $X^*$, $Y^*$ and $Z^*$ are obtained by subtracting the equilibrium solution~(\ref{34a})-(\ref{34c}) from (\ref{33a})-(\ref{33c}). Let us work with dimensionless quantities $A'$ and $\Gamma'$ defined by $A = k A'$ and $\Gamma = \gamma \Gamma'$. We also define $d$ as the diagonal  matrix with elements $d_{11}=1 $ and $d_{LL}=-1$. As a result we obtain the set of equations
 \begin{widetext}
 \begin{IEEEeqnarray}{rCl}
\label{35a}\varepsilon \nu(A'Z^*-Z^*A')+(\Gamma' Y^*+ Y^*\Gamma') + \frac{2}{\varepsilon} (Y^*-\bar{Y}^*) &=& d, \IEEEeqnarraynumspace\\[0.2cm]
\label{35b}\nu (A' X^* -X^* A') -\varepsilon (Z^*\Gamma' + \Gamma' Z^*) &=&  2 Z^*, \IEEEeqnarraynumspace\\[0.2cm]
\label{35}\nu (A' X^*+X^* A') + \varepsilon (Z^*\Gamma' - \Gamma' Z^*) &=&2Y^*,\IEEEeqnarraynumspace
\end{IEEEeqnarray}
\end{widetext}
where $\nu = k/\gamma^2$ and $\varepsilon = \gamma/\lambda$ are now the only two free dimensionless parameters. As before,  $\bar{Y}^*$ is the diagonal matrix formed by the diagonal
elements of $Y^*$.  These equations do not
involve neither $T_A$ nor $T_B$ which shows 
that $X^*$, $Y^*$ and $Z^*$ do not depend on temperature. Now,
from equation (\ref{heat_flux1}) and from the definition of
the covariance $Z$ we see that the heat flux is $J=2kZ_{n,n+1} =2k Z^*_{n,n+1}\Delta T/\lambda$ from which we may write
the following relation for the heat conductivity
\begin{equation}
\kappa = \frac{2k L Z^*_{n,n+1}}{\lambda}.
\label{37}
\end{equation}
Since $Z^*$ does not depend on temperature we conclude that
the heat conductivity does not depend on temperature.
This result is valid for any harmonic potential 
and is a direct consequence of the linearity
of the equations for the covariances \cite{rieder67}.

It is worth mentioning an important property concerning 
the position-velocity covariances. If we consider the diagonal
elements of the left and right-hand sides of equation (\ref{35a})
we get the following result
\begin{equation}
Z^*_{12}=Z^*_{23}=\ldots=Z^*_{L-1,L}
\label{45}
\end{equation}
which reflects the invariance of the heat flux along the chain
and shows that $\kappa$, given by (\ref{37}), does not depend on $n$,
as it should. It also reflects the conservation of energy inside the chain.
Incidentally, in the original harmonic chain \cite{rieder67}, which is obtained from our model by setting $\lambda = 0$, the matrix $Z$ is Toeplitz and Eq.~(\ref{45}) is thus fulfilled. When $\lambda \neq 0$, even though the first diagonal is still constant, as in Eq.~(\ref{45}), the same is not true of the others. 

%
\subsection{Large $\lambda$ expansion}
%

 As will be shown in this section, the heat conductivity in the limit of large $\lambda$ and large $L$ is described by 
\begin{equation}
\label{eq:kappa_inf}
\kappa = \frac{k L}{\frac{k}{\gamma} + c \gamma +   \lambda L}, 
\end{equation}
where $c$ is found numerically to be $c = 1.20938909(5)$. 
We call the attention to the fact that, in the thermodynamic limit, $\kappa = k/\lambda$ and the heat conductivity is thus independent of the coupling constant $\gamma$.
Formula (\ref{eq:kappa_inf}) is depicted by the upper continuous line in Fig.~\ref{fig:FSC}. As can be seen, it agrees quite well with the simulations when 
$\lambda L$ is large.
The agreement, as is expected, becomes worse when $\lambda L$ is small.

The purpose of this section is to derive formula~(\ref{eq:kappa_inf}) for the heat conductivity, valid for large $L$ and large $\lambda$. 
Exact expressions for the heat conductivity $\kappa$, Eq.~(\ref{37}), can be
obtained by exactly solving equations (\ref{35a})-(\ref{35}) for small chains.  As shown in the appendix, the results always have the 
same form of a ratio of polynomials in $\lambda$, in which
the numerator is a polynomial of one order less than the
denominator. The results obtained for small chains, from $L = 2$ up to $L=14$, show that when $\lambda$ is large, the heat conductivity has the form 
\begin{equation}
\label{eq:kappa_L}
\kappa = \frac{k L S_L }{\frac{k}{\gamma} S_L + \gamma C_L + \lambda L}.
\end{equation}
where $S_L$ and $C_L$ are rational numbers that depend on $L$. In the appendix we show the exact values of these numbers for $L=2$ up to $L=5$. 
Next we shall show that this formula is in fact valid for any $L$ and that $S_L$ and $C_L$ approach finite values, $S_L\to 1$ and
$C_L\to c$, when $L\to\infty$, thus  recovering Eq.~(\ref{eq:kappa_inf}).

We start by considering the solution of equations (\ref{35a})-(\ref{35}) for large $\lambda$ or, what is equivalent, small $\varepsilon$. 
We shall therefore assume that $X^*$, $Y^*$ and $Z^*$ can be written as a series expansion in $\varepsilon$ of the form 
\begin{IEEEeqnarray}{rCl}
X^* &=& X^0 + \varepsilon X^I + \varepsilon^2 X^{II} + \ldots, \\[0.2cm]
Y^* &=& Y^0 + \varepsilon Y^{I} + \varepsilon^2 Y^{II} + \ldots, \\[0.2cm]
Z^* &=& Z^0 + \varepsilon Z^{I} + \varepsilon^2 Z^{II} + \ldots,
\label{eq:z_series}
\end{IEEEeqnarray}
Since $\kappa$ is given by Eq.~(\ref{37}), we may also write
\begin{equation}\label{eq:kappa_series}
\kappa = \kappa^I \varepsilon  + \kappa^{II} \varepsilon^2 +\ldots,
\end{equation}
where
\begin{equation}\label{eq:kappa12}
\kappa^I =  \frac{2 k L }{\gamma}Z^0_{n,n+1}, \qquad \kappa^{II} = \frac{2 k L}{\gamma} Z^{I}_{n,n+1}
\end{equation}
Thus, our goal now is to find the functions $Z^0_{n,n+1}$ and $Z^I_{n,n+1}$.

Let us write down the ensuing equations  for each order of $\varepsilon$ that stem from Eqs.~(\ref{35a})-(\ref{35}). In order $1/\varepsilon$ the only contribution is found in Eq.~(\ref{35a}) and gives
\begin{equation}
Y^0 = \bar{Y}^0,
\end{equation}
i.e., $Y^0$ is diagonal. To order zero in $\varepsilon$ we find the following system of equations:
\begin{IEEEeqnarray}{rCl}
\label{eq:order0a} (\Gamma' Y^0 + Y^0 \Gamma') + 2 (Y^{I} - \bar{Y}^1) &=& d, \\[0.2cm]
\label{eq:order0b} \nu(A' X^0 - X^0 A') &=& 2 Z^0,  \\[0.2cm]
\label{eq:order0c} \nu(A' X^0 + X^0 A') &=& 2Y^0
\end{IEEEeqnarray}
From Eq.~(\ref{eq:order0a}) we may reach two conclusions. First, by looking at the diagonal entries we find that 
\begin{equation}\label{eq:y0meio}
Y^0_{11} = - Y^0_{LL} = 1/2.
\end{equation}
Second, since the right-hand side is diagonal, we find that
\begin{equation}
Y^{I} = \bar{Y}^I, 
\end{equation}
i.e., $Y^{I}$ is also diagonal ($Y^{II}$ will no longer be diagonal so $Y^*$, itself, is \emph{not} diagonal). 

We may now use Eqs.~(\ref{eq:order0b}) and (\ref{eq:order0c}) to eliminate $X^0$. The result is 
\begin{equation}
\label{eq:z0y0}
A' Z^0 + Z^0 A' = A' Y^0 - Y^0 A'.
\end{equation}
This matrix equation should be solved subject to the  constraint~(\ref{45}) and the boundary condition (\ref{eq:y0meio}).  
It is equivalent to $L(L-1)/2$ linear equations. Taking into account Eq.~(\ref{45}), there are $(L^2-3L+4)/2$ unknown  entries for $Z^0$. Similarly, taking into account Eq.~(\ref{eq:y0meio}), there are $L-2$ unknown entries for $Y^0$. Hence, the number of equations is the same as the number of unknowns.

\begin{figure}[!h]
\centering
\includegraphics[width=0.45\textwidth]{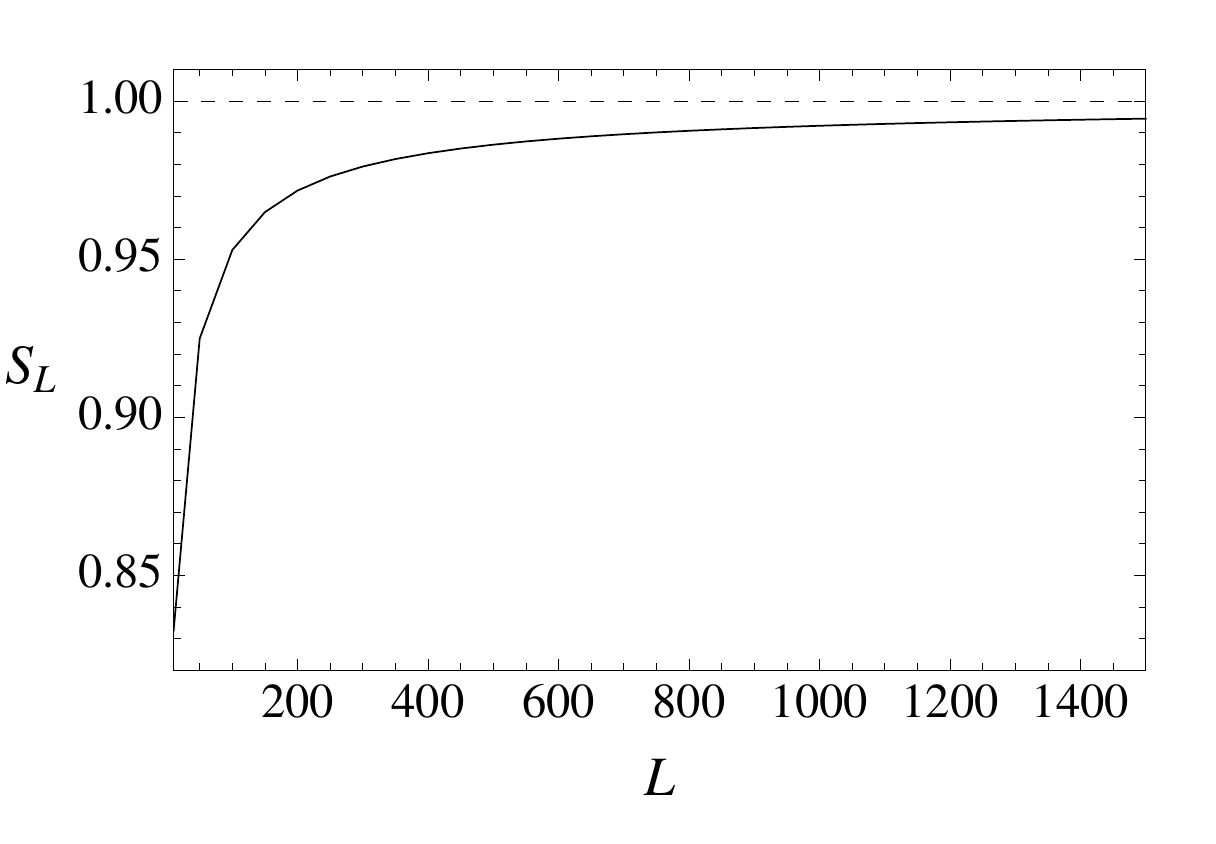}
\caption{\label{fig:S_L}The function $S_L = 2 L Z^0_{n,n+1}$, where $Z^0$ is the solution of Eq.~(\ref{eq:z0y0}).}
\end{figure}

Equation (\ref{eq:z0y0}) yields $Z^0_{n,n+1}$, from which we may obtain $\kappa^I$ by the use of  Eq.~(\ref{eq:kappa12}).
If we expand Eq.~(\ref{eq:kappa_L}) up to order $\epsilon$ we find the relation $\kappa^I = k S_L/\gamma$ between $S_L$ and $\kappa^I$.
Whence, 
\begin{equation}\label{eq:SL}
S_L = 2 L Z^0_{n,n+1}.
\end{equation} 
Note that, because of Eq.~(\ref{45}), $Z^0_{n,n+1}$ is independent of $n$, even though it depends on $L$.
The dependence of $S_L$ on $L$ is obtained by numerically solving Eq.~(\ref{eq:z0y0}) for $Z^0_{n,n+1}$.  The result is shown in Fig.~\ref{fig:S_L}. As can be seen, it approaches monotonically  the value 1. In fact, from our numerical results, $S_L -1 \sim (\ln L)/L$ when $L\to\infty$. 

We now analyze the next term in the series expansion in order to obtain  $\kappa^{II}$ in Eq.~(\ref{eq:kappa12}). The terms of order $\varepsilon$ in Eqs.~(\ref{35a})-(\ref{35}) give rise to the following system of equations 
\begin{IEEEeqnarray}{rCl}
\label{eq:order1a}(\Gamma' Y^{I} + Y^{I} \Gamma') + 2 (Y^{II} - \bar{Y}^{II} ) &=& - \nu (A' Z^0 - Z^0 A'),  \IEEEeqnarraynumspace\\[0.2cm]
\label{eq:order1b}\nu(A' X^I - X^I A') - 2 Z^{I} &=&  (Z^0 \Gamma' + \Gamma' Z^0), \IEEEeqnarraynumspace \\[0.2cm]
\label{eq:order1c}\nu(A' X^I + X^I A')-  2 Y^{I} &=&  - (Z^0 \Gamma' - \Gamma' Z^0).\IEEEeqnarraynumspace
\end{IEEEeqnarray}
From Eq.~(\ref{eq:order1a}) we conclude  (as just mentioned), that $Y^{II}$ is not diagonal. Moreover, from the first and last diagonal entries of this equation it follows that 
\begin{equation}\label{eq:y1}
Y^{I}_{11} = - Y^{I}_{LL} = - \nu Z^0_{n,n+1},
\end{equation} 
which will again serve as a boundary condition. 

Eliminating $X^I$ in Eqs.~(\ref{eq:order1b}) and (\ref{eq:order1c}) we find that 
\begin{equation}\label{eq:order1d}
(A' Z^{I} + Z^{I} A') = (A' Y^{I} - Y^{I} A') - (A'Z^0 \Gamma' + \Gamma' Z^0 A').
\end{equation}
Since the solutions are linear, we may separate $Y^{I}$ and $Z^{I}$ in two parts as 
\begin{IEEEeqnarray}{rCl}
Z^{I} &=& Z' \nu + Z'', \\[0.2cm]
Y^{I} &=& Y' \nu + Y''.
\end{IEEEeqnarray}
From Eq.~(\ref{eq:y1}) we then have that $Y'_{11} = -Y'_{LL} = - Z^0_{n,n+1}$ and $Y''_{11} = Y''_{LL} = 0$. Separating Eq.~(\ref{eq:order1d}) in two parts we find 
\begin{IEEEeqnarray}{rCl}
\label{eq:order11}A'Z' + Z' A' &=& A' Y' - Y' A', \\[0.2cm]
\label{eq:order12}A'Z'' + Z'' A' &=& A' Y'' - Y'' A' - (A'Z^0 \Gamma' + \Gamma' Z^0 A').\IEEEeqnarraynumspace
\end{IEEEeqnarray}

\begin{figure}[!h]
\centering
\includegraphics[width=0.45\textwidth]{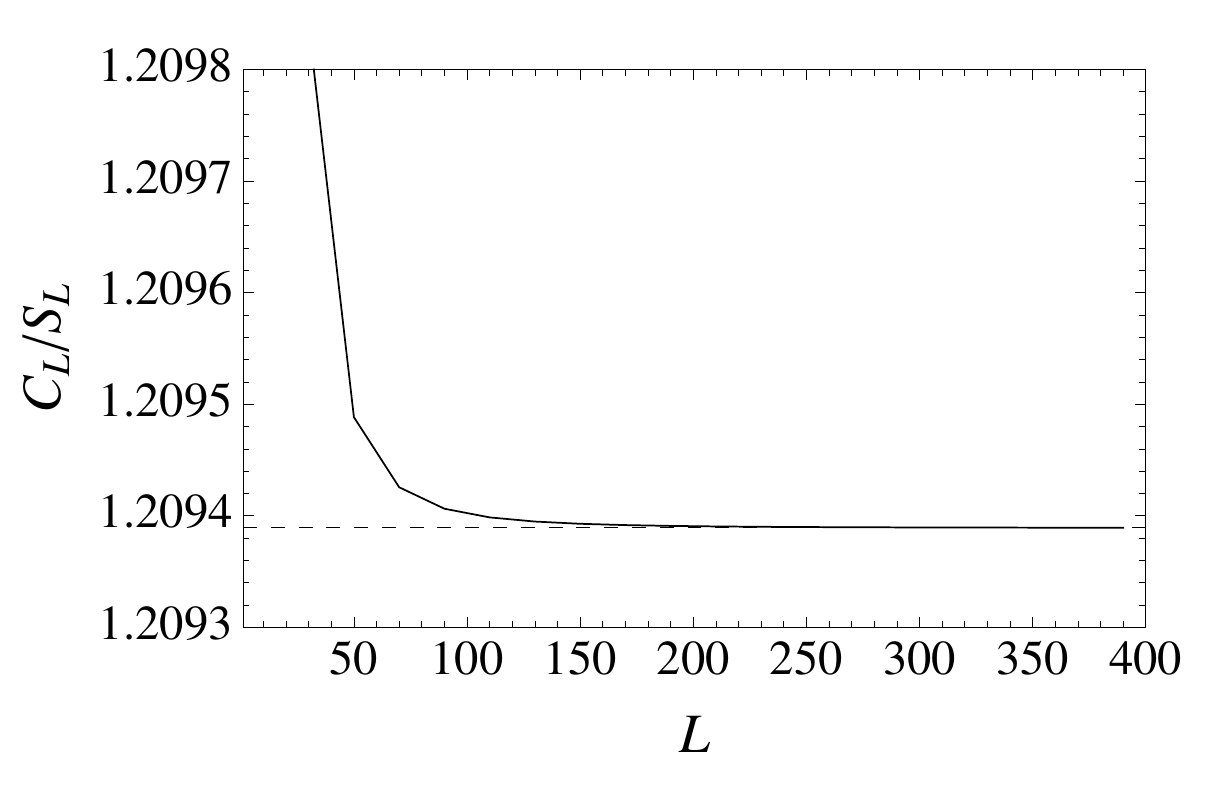}
\caption{\label{fig:C_L}The ratio $C_L/S_L$, where $C_L$ is given by Eq.~(\ref{eq:CL}) and $Z''$ is the solution of Eq.~(\ref{eq:order12}). $S_L$ is shown in Fig~\ref{fig:S_L}.}
\end{figure}

Now let us analyse our result. Referring back to Eq.~(\ref{eq:kappa12}) for $\kappa^{II}$, we may write
\begin{equation}\label{eq:b2termos}
\kappa^{II} = \frac{2 k L}{\gamma} \big(\nu Z'_{n,n+1} + Z''_{n,n+1}\big).
\end{equation}
According to Eq.~(\ref{eq:order11}), $Z'_{n,n+1}$ is given by the same equation as $Z^0_{n,n+1}$ [Eq.~(\ref{eq:z0y0})], but with the boundary condition $Y'_{11} = -Y'_{LL}=-Z^0_{n,n+1}$ instead of $Y^0_{11} =-Y^0_{LL}= 1/2$. Hence, by linearity
\begin{equation}
Z'_{n,n+1} = -2 \Big(Z^0_{n,n+1}\Big)^2.
\end{equation}
Eq.~(\ref{eq:kappa_series}) is, up to order $1/\lambda^2$, equivalent to 
\begin{equation}
\kappa = \frac{\kappa^I \gamma L}{-\frac{\kappa^{II}}{\kappa^I} \gamma L + \lambda L},
\end{equation}
or, what is equivalent, 
\begin{equation}
\kappa = \frac{ k L (2 L Z^0_{n,n+1})}{\frac{k}{\gamma} (2 L Z^0_{n,n+1}) + \gamma \left(-\frac{L Z''_{n,n+1}}{Z^0_{n,n+1}}\right) + \lambda L}.
\end{equation}
By comparing this result with Eq.~(\ref{eq:kappa_L}) it is clear that  $S_L = 2 L Z^0_{n,n+1}$ and 
\begin{equation}\label{eq:CL}
C_L = - \frac{L Z''_{n,n+1}}{Z^0_{n,n+1}}.
\end{equation}
The dependence of $C_L$  on $L$ is obtained by numerically solving Eq.~(\ref{eq:order12}), using $Z^0$, previously obtained, as input. In Fig.~\ref{fig:C_L} we show the result for $C_L/S_L$, since it converges much faster with $L$. The asymptotic value $C_\infty = c$ is found to be
\begin{equation}\label{eq:C_infty}
c =  1.20938909(5).
\end{equation}

To summarise the results of this section, we have shown that, for large values of $\lambda$,
the heat conductivity behaves according to (\ref{eq:kappa_L})
which for sufficient large $L$ reduces to the expression (\ref{eq:kappa_inf}) or, what is equivalent, Eq.~(\ref{eq:kappa_primeira}).

\subsection{Fourier method of computing $S_L$}

\begin{figure}[!t]
\centering
\includegraphics[width=0.5\textwidth]{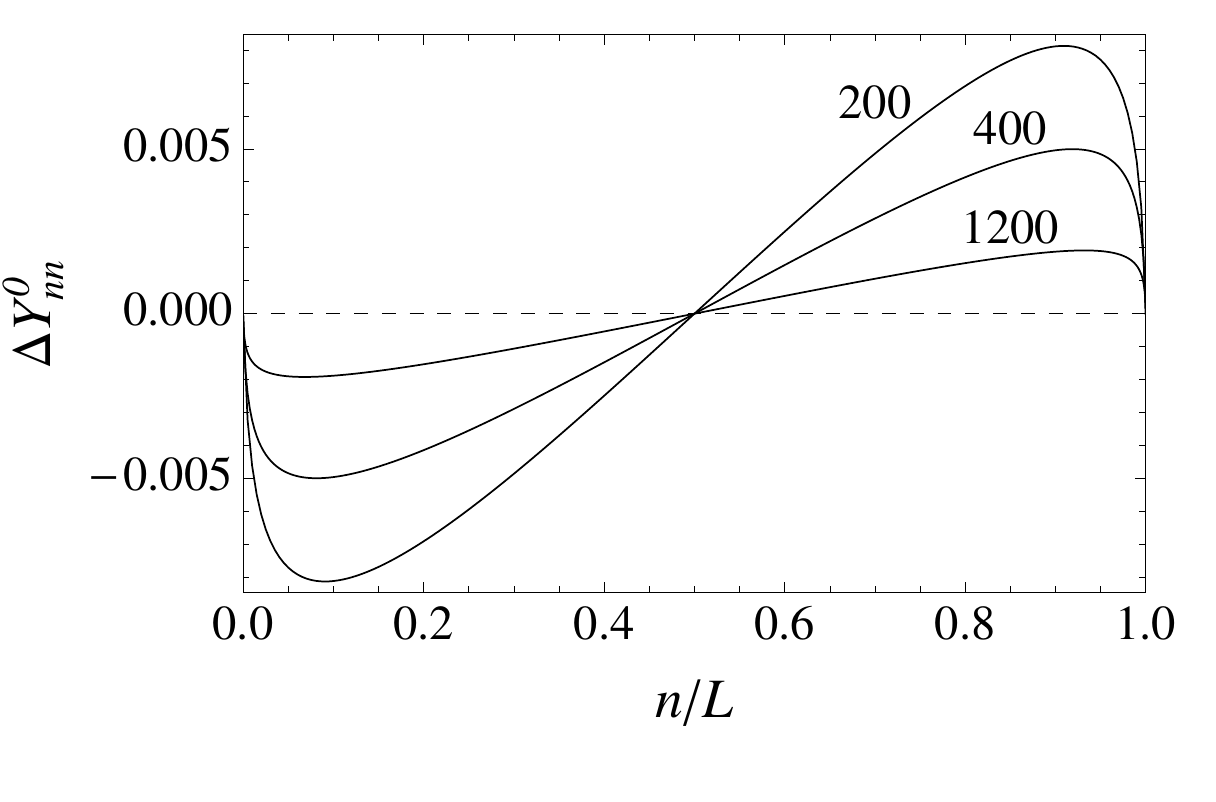}
\caption{\label{fig:Y}Difference between the exact numerical solution for $Y^0_{nn}$ [Eq.~(\ref{eq:z0y0})] and the linear profile given by  the right-hand side of~(\ref{41}) for different system sizes $L$, as indicated.}
\end{figure}

We now illustrate how to obtain the function $S_L$ analytically by a different approach. 
Our goal is again to solve Eq.~(\ref{eq:z0y0}) with $Y^0_{11} = - Y^0_{LL} = 1/2$. The solution will be based on the assumption that, for large $L$, the diagonal matrix $Y^0$ approaches a linear profile between  $1/2$ and $-1/2$. 
This fact can be verified from the numerical solution of Eq.~(\ref{eq:z0y0}), as illustrated in Fig.~\ref{fig:Y}, which shows the difference $\Delta Y^0_{nn}$ between the exact numerical solution and the linear interpolation. As can be seen in the Fig.~\ref{fig:Y}, this difference vanishes in the limit $L\to\infty$. 
This assumption is also  reasonable given that the diagonal entries of $Y$ represent the mean-squared velocity profile, which should be linear if the system is to obey Fourier's law. 
Hence, we shall take 
\begin{equation}
Y^0_{nn} = h(L+1-2n),
\label{41}
\end{equation}
where $h=1/[2(L-1)]$,
which interpolates linearly between the values $Y^0_{11}=1/2$ and $Y^0_{LL}=-1/2$.

Equation (\ref{eq:z0y0}) can be solved for $Z^0$ by diagonalizing
$A$. The matrix that diagonalizes $A$ is obtained from its eigenvectors,
which are
\begin{equation}
\psi_{kn} = \sqrt{\frac{2}{L+1}}\sin kn,
\end{equation}
where $k=\pi j/(L+1)$, $j=1,2,\ldots,L$. 
Defining $\hat{Y}^0_{kq}$ and $\hat{Z}^0_{kq}$ by
\begin{equation}
\hat{Y}^0_{kq} = \sum_{nm} \psi_{kn} Y^0_{nm}\psi_{kn},
\label{42}
\end{equation}
\begin{equation}
\hat{Z}^0_{kq} = \sum_{nm} \psi_{kn} Z^0_{nm}\psi_{kn}, 
\end{equation}
where $q=\pi \ell/(L+1)$, $\ell=1,2,\ldots,L$,
we get from equation (\ref{eq:z0y0}) the following relation
between these quantities
\begin{equation}
\hat{Z}^0_{kq}=\frac{\cos q - \cos k}{2-\cos k -\cos q}\hat{Y}^0_{kq}.
\label{51}
\end{equation}

Now, replacing (\ref{41}) into (\ref{42}),  and performing the summation we get
\begin{equation}
\hat{Y}^0_{kq} = \frac{-4h\sin k\sin q}{(L+1)(\cos k -\cos q)^2},
\label{42c}
\end{equation}
valid for $j+\ell$ odd. When $j+\ell$ is even, the summation
vanishes and $\hat{Y}^0_{kq} =0$. 

To get $Z^0_{nm}$ from $\hat{Z}^0_{kq}$, we use the inverse transformation
\begin{equation}
Z^0_{nm} = \sum_{kq} \psi_{kn} \hat{Z}^0_{kq}\psi_{km}. 
\label{53}
\end{equation}
Inserting (\ref{42c}) and (\ref{51}) into (\ref{53}) gives 
\begin{equation}
Z^0_{nm} = \frac{8h}{(L+1)^2}\sum_{kq} 
\frac{\sin kn \sin q m}{2-\cos k -\cos q}\,
\frac{\sin k\sin q}{\cos k-\cos q},
\label{54}
\end{equation}
where the summation is over $j+\ell$ odd. This sum may be computed numerically for large enough $L$ and $m = n+1$. 

\begin{figure}[!t]
\centering
\includegraphics[width=0.45\textwidth]{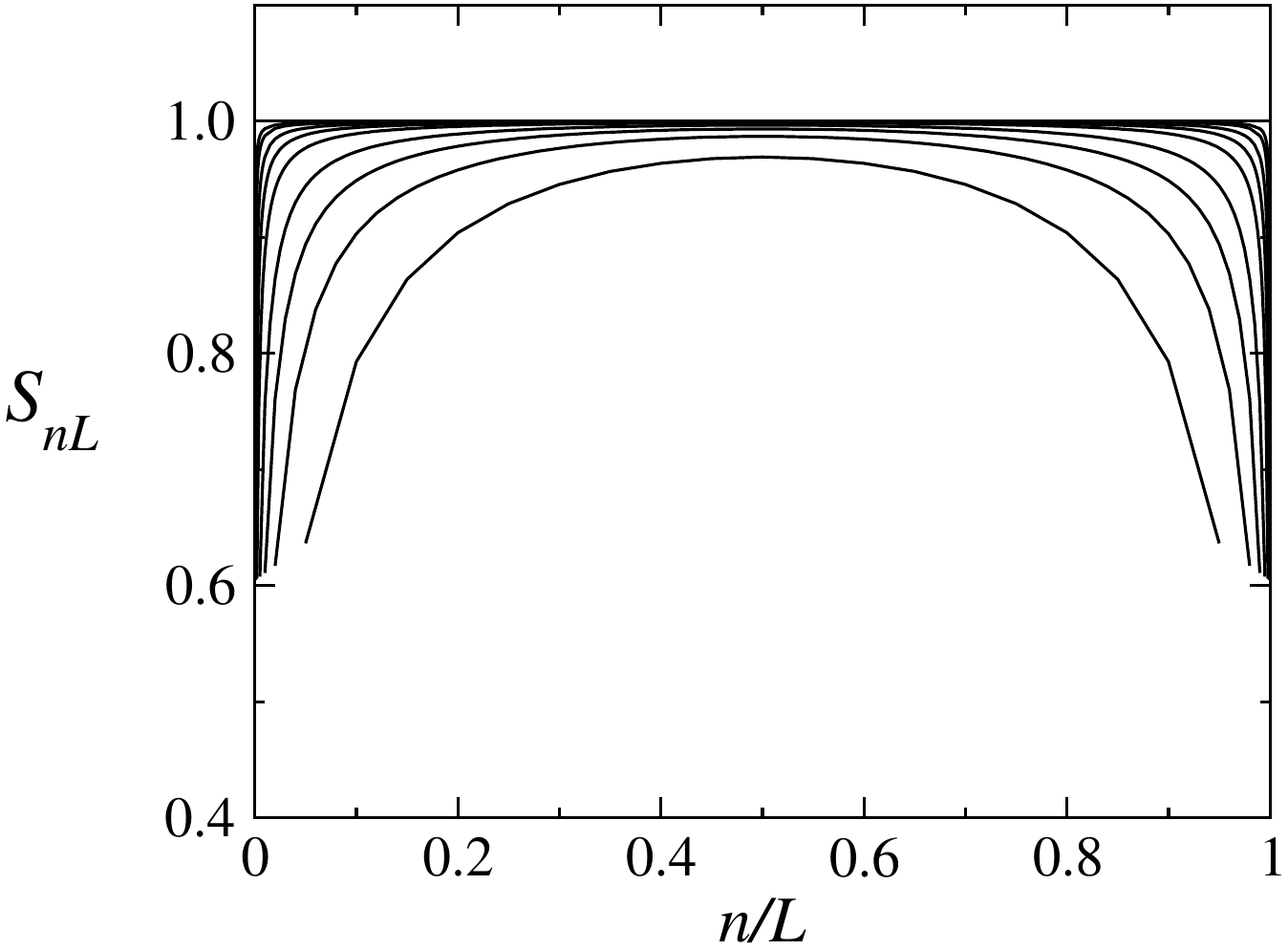}
\caption{\label{fig:z}Numerical calculation of $S_{nL} = 2L Z^0_{n,n+1}$, where $Z^0_{n,n+1}$ is given by Eq.~(\ref{54}). The curves are for different values of $L$, from  bottom to top, starting at $L =20$ and going up to $L = 10000$.}
\end{figure}

The function $S_{nL} = 2 L Z^0_{n,n+1}$, computed numerically from Eq.~(\ref{54}), is shown in Fig.~\ref{fig:z} for several values of $L$. As can be seen, the results depend on $n$, a consequence of the linear interpolation approximation~(\ref{41}). However,
When $L\to\infty$, $S_{nL}$ approach a constant value, namely the value
one, as seen in Fig.~\ref{fig:z}, further corroborating the results of Fig~\ref{fig:S_L} for $S_L$, when $L\to\infty$,
which was precisely the purpose of this calculation. 

%
%
\section{Discussions and Conclusions}
%
%

As our last topic, let us briefly digress about the form of the probability distribution in the NESS. The general solution of the Fokker-Planck equation~(\ref{9}) for $P(x,v,y,u)$ is, \emph{in the steady state}, given by a multivariate Gaussian distribution. This can be seen by direct substitution in the
Fokker-Planck equation (\ref{9}),
which is simpler to do if one uses the characteristic function $G(k',k'',q',q'')$ defined as 
\begin{equation}
G = \int e^{i (k'\cdot x + k'' \cdot v + q' \cdot y  + q'' \cdot u)} P dx dv dy du.
\end{equation}
It is then possible to verify explicitly that the steady-state solution for $G$ is given by 
\begin{equation}
G = \exp{\left\{ -\frac{1}{2} \Big(k^\dagger \Theta_1 k + q^\dagger \Theta_2 q + 2 k^\dagger \Theta_3 q\Big)\right\}},
\end{equation}
where $k = (k',k'')$, $q = (q',q'')$ and the covariance matrices $\Theta_1$, $\Theta_2$ and $\Theta_3$ are defined in Eq.~(\ref{20}). The system is therefore described entirely by the covariance matrices, whose entries have been determined in the previous sections. It is worth mentioning that in the particular case of the potential $U_1$ in Eq.~(\ref{14}), the matrix $\Theta_3$ is identically zero and, moreover, $\Theta_1 = \Theta_2$; i.e., the variables $(x,v)$ and $(y,u)$ become statistically independent. Finally, we note that the time-dependent solution is not necessarily given by a multivariate Gaussian. However, if the system starts with a Gaussian distribution, it remains Gaussian indefinitely.  

In conclusion, we have introduced a modification of the harmonic chain whereby all particles are also subject to elastic collisions that conserve the kinetic energy. As was shown, it reproduces Fourier's law irrespective of the intensity of the collisions. These results corroborates our argument that the fine details of the noise are unimportant in leading to Fourier's law; but, rather, that what is relevant is its  energy-conserving nature. 
The model was solved using a numerically exact procedure which is extremely efficient computationally and is valid for any type of harmonic interaction potential. For a particular choice of the interaction potential, 
we have determined the heat conductivity exactly for small chains
and also by an expansion in $\lambda^{-1}$. The first term in the
expansion were also determined by an approximation that becomes exact in the thermodynamic limit providing the exact expression $\kappa=k/\lambda$
for the heat conductivity for large enough $\lambda$ in the 
thermodynamic limit.

\section*{Acknowledgment}

We acknowledge the Brazilian agencies FAPESP and CNPq
for financial support.


\appendix
%
%
\section{Exact solution for small systems}
%
%

Closed forms for the heat conductivity of small chains
can be determined by solving the equations for the covariances.
This was accomplished using symbolic computing to solve Eq.~(\ref{30}), which is  valid specifically for the potential $U_1$ in Eq.~(\ref{14}).  We were able to find the solutions up to $L=14$, in which case there were more than 300 coupled linear equations (hence the need for symbolic computing). The results always have the form of a ratio of polynomials in $\lambda$, viz., 
\begin{equation}\label{eq:poly}
\kappa_L = \frac{\displaystyle{\sum\limits_{j = 0}^M p_j \lambda^j}}{\displaystyle{\sum\limits_{j = 0}^{M+1} q_j \lambda^j}}
\end{equation}
The degree of the polynomial in the numerator is  $M$ and that of the denominator, $M+1$, where $M$ turns out to be 
\begin{equation}
M = \begin{cases}
\frac{L^2}{2}-L			& \text{if } L \text{ is even,} \\[0.2cm]
\frac{L^2}{2}-L+\frac{1}{2}	& \text{if } L \text{ is odd.} 
\end{cases}
\end{equation}

For the purpose of illustration, we show the results from $L=2$ to $L=4$:
\begin{widetext}
\begin{IEEEeqnarray*}{rCl}
\kappa_2 &=& \frac{2k }{\frac{k}{\gamma}+2 \gamma + 2  \lambda} ,\\[0.2cm]
\kappa_3 &=& \frac{3k (k + 2 \gamma^2 + 6 \gamma \lambda + 4 \lambda^2)}{(\frac{k^2}{\gamma} + 4 k\gamma + 3 \gamma^3) + (10 k+16\gamma^2)\lambda + (4 \frac{k}{\gamma} + 27 \gamma) \lambda^2 + 14 \lambda^3},\\[0.2cm]
\kappa_4 &=& \frac
{4k[(k^2 + 4 k \gamma^2 + 3 \gamma^4) + (14 k \gamma + 22 \gamma^3) \lambda + (12 k + 59 \gamma^2)\lambda^2 + 68 \gamma \lambda^3 + 28 \lambda^4]}
{q_0+ 4(5 k^2 + 17 k \gamma^2 + 11 \gamma^4) \lambda + (12\frac{k^2}{\gamma} + 155 k \gamma + 186 \gamma^3)\lambda^2+
6(22k + 63 \gamma^2)\lambda^3 + 4(7\frac{k}{\gamma} + 92 \gamma) \lambda^4 + 136 \lambda^5},
 \end{IEEEeqnarray*}
 \end{widetext}
where $q_0 = (\frac{k^3}{\gamma} + 6 k^2 \gamma + 10 k \gamma^3 + 4 \gamma^5)$.

Retaining the dominant terms in $\lambda$ in the numerator and denominator
we may cast them in the form (\ref{eq:kappa_L}). This is tantamount to determining exactly the functions $S_L$ and $C_L$ in Eq.~(\ref{eq:kappa_L})  for small values of $L$. The results for $L = 3$ and $L=4$ are ($\kappa_2$ is already in the form (\ref{eq:kappa_L})):
\begin{IEEEeqnarray*}{rCl}
\kappa_3 &=& \frac{\frac{6}{7} (3k)}{\frac{6}{7} \frac{k}{\gamma} + \frac{9}{7} \gamma + 3 \lambda},\\[0.2cm]
\kappa_4 &=& \frac{\frac{14}{17} (4 k)}{\frac{14}{17} \frac{k}{\gamma} + \frac{132}{119} \gamma + 4 \lambda},\\[0.2cm]
\kappa_5 &=& \frac{\frac{22}{27} (5k)}{\frac{22}{27}\frac{k}{\gamma} + \frac{311}{297} \gamma + 5 \lambda},\\[0.2cm]
\kappa_6 &=& \frac{\frac{1485}{1823} (6k)}{\frac{1485}{1823}\frac{k}{\gamma} + \frac{307618}{300795} \gamma + 6 \lambda}.
\end{IEEEeqnarray*}

\end{document}